\documentclass[aps,prl,twocolumn]{revtex4-1}
\usepackage[utf8]{inputenc}
\usepackage{bm}
\usepackage{amssymb}
\usepackage{mathtools}
\usepackage{amsmath}
\usepackage{xcolor}
\usepackage{enumitem}
\usepackage{natbib}
\usepackage{hyperref}
\usepackage{xcolor}
\usepackage{xcolor,colortbl}

\usepackage{diagbox}
\usepackage[ruled,vlined]{algorithm2e}

\begin{document}

\title{Layered Complex Networks as Fluctuation Amplifiers}
\author{Melvyn Tyloo}
\affiliation{Theoretical Division, Los Alamos National Laboratory, Los Alamos, NM 87545 USA.}
\date{\today}

\begin{abstract}
	In complex networked systems theory, an important question is how to evaluate the system robustness to external perturbations. With this task in mind, I investigate the propagation of noise in multi-layer networked systems. I find that, for a two layer network, noise originally injected in one layer can be strongly amplified in the other layer, depending on how well-connected are the complex networks in each layer and on how much the eigenmodes of their Laplacian matrices overlap. These results allow to predict potentially harmful conditions for the system and its sub-networks, where the level of fluctuations is important, and how to avoid them. The analytical results are illustrated numerically on various synthetic networks.
\end{abstract}

\maketitle

\begin{figure*}
    \centering
    \includegraphics[scale=0.4]{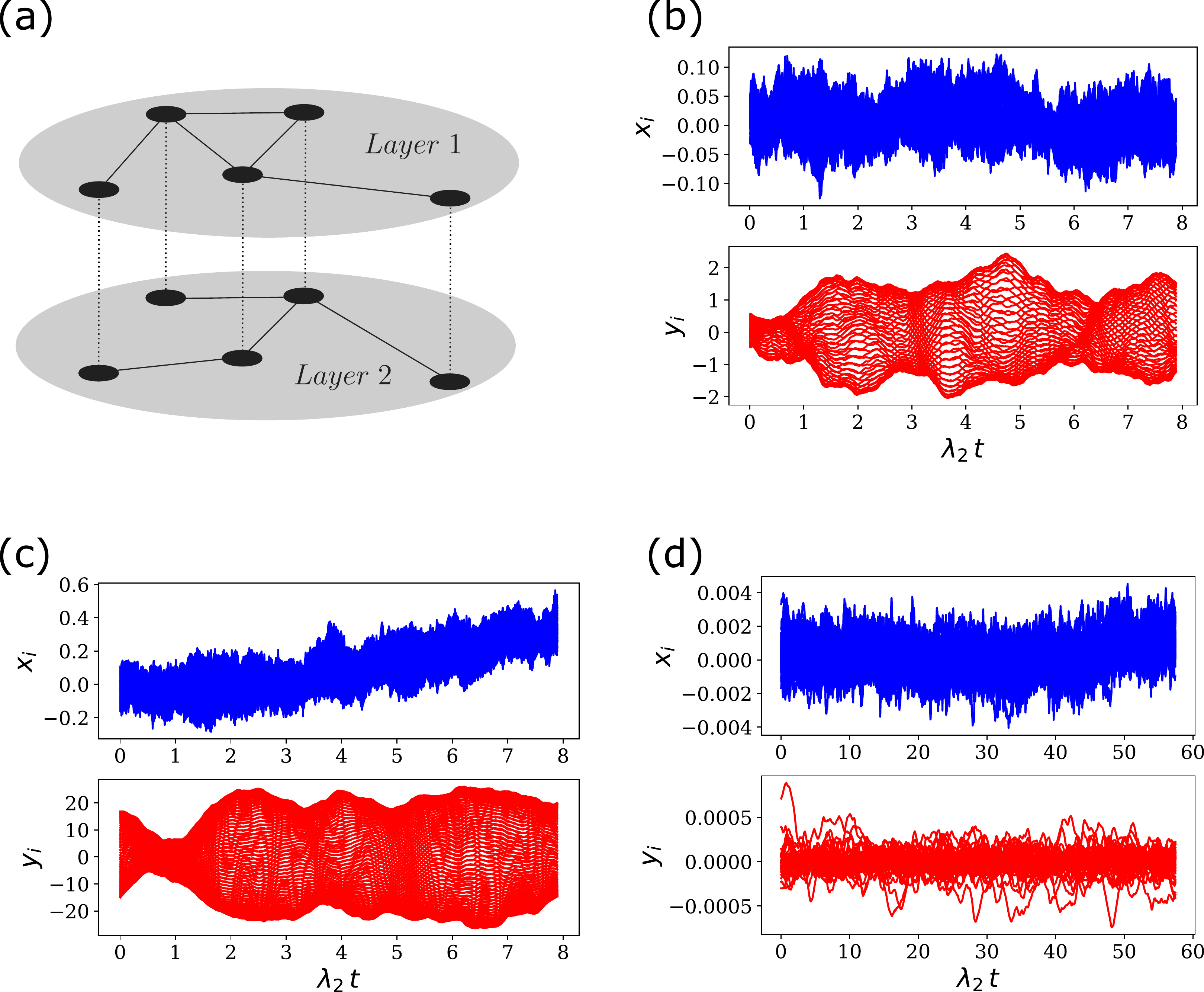}
    \caption{(a) Illustration of a two-layer system. Each grey area corresponds to a layer with its own coupling network. The two layers are interacting through some coupling function represented by dotted lines. (b)-(d) Trajectories $x_i$ and $y_i$ respectively in layer 1 and 2 for the same network in both layers: (a) cycle networks of size $n=50$, (b) cycle networks of size $n=100$, (d) Erd\H{o}s-R\'enyi networks of size $n=50$ with edge probability $p=0.15$. For readability, trajectories are only shown for a limited time interval. Amplification of the fluctuations for cycle networks with (b) $\lambda_2\cong 0.0158$ and (c) $\lambda_2\cong 0.004$\,. (d) Reduction of the fluctuations for Erd\H{o}s-R\'enyi networks $\lambda_2\cong 2.87$\,. For each panel, the mean amplification factor is given by $\frac{\sum_i\langle y_i^2\rangle}{\sum_i\langle x_i^2 \rangle} = 1209$ (b), $20181$ (c), $0.02$ (d). }
    \label{fig1}
\end{figure*}

\textbf{Introduction}.-- Complex-networked systems are widely present in physical and man-made applications ranging from neurons in the brain to large-scale power transmission networks~\cite{Bar16}. These systems are made of individual elements with their own internal dynamics, that are interacting together. The interplay between the internal degrees of freedom and the coupling typically gives birth to organized collective dynamics such as consensus or synchronization~\cite{Str04,Pik03}. The coupling between the elements is usually modelled by complex networks. Recently a lot of effort has been put in the refinement of this approach in order to account for specific structures that might exist in the coupling~\cite{Kur06,DeD13,Bia15,BiaB18}. Indeed, many complex systems are interdependent or hierarchically influence each other~\cite{Xia21}, as e.g. in the brain where various regions with different functional networks interact with each other~\cite{Ped18}, or in power networks where layers with different voltage levels are connected~\cite{Mac08}, or in social networks where people are part of interacting communities~\cite{Mur14}. For such applications, the extension to multi-layer networks as depicted in Fig.~\ref{fig1}(a) is particularly relevant. The latter are composed of many different layers of networks interacting with one another. The nodes in the different layers might represent the same individual element taking part in different coupled dynamics, or distinct elements belonging to separate systems that somehow interact together. Also, these systems are unavoidably subject to noise coming from the interacting elements or from the environment~\cite{Kam76}. Such noisy conditions might have different effects on the coupled system. Indeed, depending on its amplitude and time-dependence, it might induce fluctuations around a stationary state or even escapes from an initial basin of attraction~\cite{Dev12,Hin16,Tyl18c}. Both outcomes are potentially harmful to the system as fluctuations might damage some network components on the long run, and basin escapes can lead to large system failures. Therefore, to prevent such events and in order to devise more resilient interdependent complex systems~\cite{Ron18,Nag22}, it is of particular interest to understand how noise in one layer impact the dynamics of other layers.

In this letter, I consider two-layer systems where only one layer is subjected to noise. Interestingly, I find that, depending on the smallest eigenvalues of the sub-network Laplacian matrices and the overlap of their eigenmodes, fluctuations can be significantly amplified or reduced in the second layer. I show the latter analytically by comparing the second moment of the degrees of freedom in each layer. I also illustrate the theory numerically on different networks and extend the principle to networks with more than two layers.\\

\textbf{Two layers networked system}.-- I consider two layers of diffusively coupled agents, $\{x_k\}$, $\{y_k\}$. Each layer has $n$ nodes and its own undirected complex coupling network. The two layers are coupled together in a directed way that is detailed later. The dynamics of the $2n$ agents is governed by the following set of differential equations,
\begin{eqnarray}
\dot{x}_i &=& - \sum_{j=1}^n \mathbb{L}_{ij}^{(1)}\,x_j + \eta_i  \quad  i=1,...n\,,\label{eq1}\\
\dot{y}_i &=& - \sum_{j=1}^n \mathbb{L}_{ij}^{(2)}\,y_j + f_i(\{x_k\},\{y_k\}) \quad  i=1,...n\,,\label{eq2}
\end{eqnarray}
where $\{x_k\},\{y_k\}$ are the nodal degrees of freedom respectively in layer 1 and 2, $\mathbb{L}_{ij}^{(l)}$ is the Laplacian matrix of the undirected coupling network in the $l$-th layer, and $f_i$ is a coupling function between the two layers. As set in Eq.~(\ref{eq2}), the inter-layer interaction is directed, i.e. Eq.~(\ref{eq1}) does not depend on $\{y_k\}$\,, while Eq.(\ref{eq2}) does depend on $\{x_k\}$. To investigate the propagation of noise from the first to the second layer, only the first one is subjected to ambient noise $\eta_i$ that is taken as white, i.e. $\langle\eta_i(t)\eta_j(t')\rangle = \delta_{ij}\,\eta_0^2\,\delta(t-t')$~\cite{rem1}. Without loss of generality, I take the initial conditions to be $x_i(0)=0$\,, $y_i(0)=0$ for $i=1,...n$\,. Indeed, in the following I consider the long time limit for the second moment which is independent of the initial conditions. Importantly, Eqs.~(\ref{eq1}), (\ref{eq2}) can represent the linear approximation of a nonlinear system around a stable fixed point.

In the following, I denote $\lambda_\alpha^{(k)}$\, , $u_{\alpha,i}^{(k)}$ the eigenvalues and corresponding eigenvectors respectively, of the Laplacian matrix $\mathbb{L}^{(k)}$\,. As I consider connected undirected networks in each layer, I have $0=\lambda_1^{(k)}<\lambda_2^{(k)}\le...\le\lambda_n^{(k)}$ for $k=1,2$ with the first eigenvector being constant i.e. $\bm u_{1}^{(k)}=(1,...,1)/\sqrt{n}$\, for $k=1,2$\,. The second smallest eigenvalue, namely $\lambda_2^{(k)}$ is usually called \textit{algebraic connectivity} of the network, with its corresponding eigenvector ${\bm u}_2^{(k)}$ called \textit{Fiedler mode}.\\

\textbf{Noise propagation}.-- First, one has to choose an inter-layer coupling function. The simplest choice is  $f_i(\{x_k\},\{y_k\}) =  x_i - n^{-1}\sum_j x_j$\,, i.e. the first layer tunes the natural velocities of the second one, orthogonal to the zero mode $\bm u_1^{(2)}$~\cite{rem2}. Using this coupling function together with a modal decomposition over the eigenvectors of $\mathbb{L}^{(1)}$ and $\mathbb{L}^{(2)}$\,, one obtains as solution of Eqs.~(\ref{eq1}),(\ref{eq2}) (see \cite{Tyl18a, Tyl19} for details),
\begin{eqnarray}\label{eq3}
x_i(t) &=& \sum_\alpha e^{-\lambda_\alpha^{(1)}t} \int_0^t e^{\lambda_\alpha^{(1)}t'}\,\sum_j \eta_j\, u_{\alpha,j}^{(1)}\, {\mathrm{d}t'}\,u_{\alpha,i}^{(1)}\,, \\
y_i(t) &=& \sum_\alpha e^{-\lambda_\alpha^{(2)}t} \int_0^t e^{\lambda_\alpha^{(2)}t'}\,\sum_j x_j\, u_{\alpha,j}^{(2)}\, {\mathrm{d}t'}\,u_{\alpha,i}^{(2)}\,.\label{eq32}
\end{eqnarray}
Therefore, the response of the first layer depends on the scalar product between ${\bm \eta}$ and $\bm u_{j}^{(1)}$\,, while in the second layer it is between ${\bm x}$ and $\bm u_{j}^{(2)}$\,.
In order to investigate how noise is transmitted from one layer to the other, one may calculate the second moment of the degrees of freedom, namely $\{x_i\}$ and $\{y_i\}$. Using Eq.~(\ref{eq3}) one has in the long time limit for the first layer~\cite{Tyl19},
\begin{eqnarray}\label{eq4}
\langle x_i^2 \rangle &=& \frac{\eta_0^2}{2}\sum_\alpha \frac{{u_{\alpha,i}^{(1)}}^2}{\lambda_\alpha^{(1)}}\,,
\end{eqnarray}
which can be related to the inverse of the resistance centrality of the $i-$th node~\cite{Kle93,Tyl19}. This means the highest resistance centrality, the least fluctuations. In the model considered here, the $x_i$'s modify the natural velocities in the second layer. In order to calculate the second moment in the second layer, one needs the two-point time shifted correlators of the $x_i$'s. Due to the coupling in the first layer, the latter is then non-trivial and reads,
\begin{eqnarray}\label{eq42}
\langle x_i(t) x_j(t') \rangle = \frac{\eta_0^2}{2}\sum_\alpha \frac{{u_{\alpha,i}^{(1)}}u_{\alpha,j}^{(1)}}{\lambda_\alpha^{(1)}}e^{-\lambda_\alpha^{(1)}|t-t'|}.
\end{eqnarray}
Therefore, for the second layer, the input is a time- and space-correlated noise. For equal times, i.e. $t=t'$\,, one recovers the pseudo-inverse of $\mathbb{L}^{(1)}$~\cite{Ren10}.
Then, plugging Eq.~(\ref{eq42}) into Eq.~(\ref{eq32}) yields for the second moment in the second layer,
\begin{widetext}
\begin{align}\label{eq5}
\begin{split}
\langle y_i^2 \rangle = \frac{\eta_0^2}{2}\sum_{\alpha, \beta, \gamma} \sum_{k,l} \frac{{u_{\gamma,k}^{(1)}}u_{\gamma,l}^{(1)} {u_{\alpha,k}^{(2)}}u_{\beta,l}^{(2)} [2\lambda_\gamma^{(1)} +\lambda_\alpha^{(2)} + \lambda_\beta^{(2)}] }{\lambda_\gamma^{(1)} (\lambda_\alpha^{(2)} + \lambda_\beta^{(2)} ) ( \lambda_\gamma^{(1)} + \lambda_\alpha^{(2)} ) ( \lambda_\gamma^{(1)} + \lambda_\beta^{(2)} )  }{u_{\alpha,i}^{(2)}}u_{\beta,i}^{(2)} \,.
\end{split}
\end{align}
\end{widetext}
The fluctuation of $y_i$ is thus a non-trivial function of the overlap between $\bm u_{\gamma}^{(1)}$ and $\bm u_{\beta}^{(2)}$ and their corresponding eigenvalues. Equation~\ref{eq5} is obtained for two different complex networks in the two layers. It is easier to understand what is going on by considering the specific case of similar layers, i.e. with the same complex network. In this particular setting, one has $\bm{u}_{\alpha}^{(1)}=\bm{u}_{\alpha}^{(2)}=\bm{u}_{\alpha}$, $\lambda_\alpha^{(1)}=\lambda_\alpha^{(2)}=\lambda_\alpha$ $\forall \alpha$\,. Then Eq.~(\ref{eq5}) simplifies to,
\begin{align}\label{eq6}
\begin{split}
\langle y_i^2 \rangle = \frac{\eta_0^2}{4}\sum_{\alpha}  \frac{u_{\alpha,i}^2}{\lambda_\alpha^3} \,.
\end{split}
\end{align}
Now to see how the noisy first layer affects the second one, it is important to compare Eq.~(\ref{eq6}) to Eq.~(\ref{eq4}). Indeed, while in Eq.~(\ref{eq4}), one has the first power of $\lambda_\alpha$ at the denominator, in Eq.~(\ref{eq6}) it is the third power of $\lambda_\alpha$\,. This seemingly tiny difference might have serious implications. In particular, the slowest modes of the coupling network, i.e. the $\lambda_\alpha$'s that are the closest to zero, primarily determine whether the noise is amplified or reduced in the second layer. Therefore, one expects poorly connected networks with low algebraic connectivity to be good candidates to amplify the fluctuations, while well-connected networks might efficiently reduce the fluctuations.\\

\textbf{Numerical Results}.-- First, I illustrate the case with two times the same coupling network in both layers, i.e. $\mathbb{L}^{(1)}=\mathbb{L}^{(2)}$\,. Figure~\ref{fig1}(b)-(d) show the trajectories of $x_i$'s and $y_i$'s respectively in layer 1 and 2. Following Eq.~(\ref{eq6}), to obtain the amplification of the fluctuations, I choose cycle networks which have an algebraic connectivity that decreases with the size, namely $\lambda_2 = 2-2\cos(2\pi/n)$\,. This is shown in panel Fig.~\ref{fig1}(b) where one clearly sees that the fluctuations in the second layer (bottom panel) are strongly amplified, compared to the input noise of the first layer (top panel). In the thermodynamic limit, the amplification factor for such systems scales with the size as $\langle y_i^2 \rangle/\langle x_i^2\rangle \propto n^4 $\,. The latter is observed by comparing panels (b) and (c) where the network size has been doubled and further confirmed by the amplification factors given in the caption of Fig.~\ref{fig1}. To obtain the other counter effect, namely the reduction of the fluctuations, one needs a network with larger algebraic connectivity. Therefore, I simulate two layers of the same Erd\H{o}s-R\'enyi network with edge probability $p=0.15$\, which has considerably larger $\lambda_2$ [see panel (d)]. The second moment in the second layer is significantly reduced compared to the first one.

Second, I consider the case of different networks in each layer. Here, going back to Eq.~(\ref{eq5}), fluctuations in the second layer crucially depend on the overlap between the Laplacian eigenvectors of the two layers. In particular, if the eigenvectors do not overlap enough, only little fluctuations will propagate in the second layer. To demonstrate that, I consider the overlap between the Fiedler modes of each layer and the amplification factor for all combinations of five coupling networks of size $n=50$\,, namely, a single cycle, two Watts-Strogatz (WS) networks with different rewiring probabilities~\cite{Wat98}, a Barab\'asi-Albert (BA) network with $m=4$~\cite{Bar16}, and an Erd\H{o}s-R\'enyi (ER) network~\cite{New18book}. This is shown in Tab.~\ref{tab1}. One observes a strong amplification for combinations with the cycle and WS I. This is due to the small algebraic connectivity in each layers and the overlap that is still significant (0.405) due to the low rewiring probability for the WS I. One also notice a strong reduction of the fluctuations for combinations of BA and ER networks which satisfy both $\lambda_2^{(1)},\lambda_2^{(1)}>1$ and $({\bm u}_2^{(1)}\cdot{\bm u}_2^{(2)})^2\ll 1$\,. Interestingly, for the networks investigated here, having BA or ER in the first layer only allows reduction of the fluctuations in the second layer, while having a cycle, WS I or WS II in the first layer can lead to both amplification or reduction, depending on the type of network in the second layer.
\\

\begin{table*}
  \centering
  \begin{tabular}{|c||c|c|c|c|c|c|c|}
    \hline
    \backslashbox{$y_i$}{$x_i$}& \shortstack{Cycle \\ $\lambda_2^{(1)}=0.0158$} &  \shortstack{WS I \\ $\lambda_2^{(1)}=0.0156$} & \shortstack{WS II \\ $\lambda_2^{(1)}=0.273$} & \shortstack{BA \\ $\lambda_2^{(1)}=2.302$} & \shortstack{ER \\ $\lambda_2^{(1)}=3.02$ } \\
    \hline\hline
        \shortstack{Cycle \\ $\lambda_2^{(2)}=0.0158$} & \shortstack{$1$ \\ $819$}  & \shortstack{$0.405$ \\ $744$}  & \shortstack{ $0.286$ \\ $23.856$}  & \shortstack{$0.131$ \\ $0.905$} & \shortstack{$2.8\times 10^{-4}$ \\ $0.766$} \\
    \hline
        \shortstack{WS I \\$\lambda_2^{(2)}=0.0156$} & \shortstack{$0.405$ \\ $725$ }&  \shortstack{$1$ \\ $673.51$} & \cellcolor{red!16} \shortstack{$2.5\times 10^{-5}$ \\ $19.81$}  & \shortstack{$0.0197$ \\  $0.881$} & \shortstack{$0.0113$ \\ $0.738$}  \\
        \hline
        \shortstack{WS II \\$\lambda_2^{(2)}=0.273$} & \shortstack{$0.286$ \\ $3.1$} &  \cellcolor{red!16} \shortstack{$2.5\times 10^{-5}$\\  $2.673$ }&  \shortstack{$1$ \\$1.346$ }& \shortstack{$0.126$ \\$0.0733$ }& \shortstack{$0.0123$ \\ $0.0535$}   \\
    \hline
    \shortstack{BA \\ $\lambda_2^{(2)}=2.302$} &  \shortstack{$0.131$ \\    $0.048$} &  \shortstack{$0.0197$ \\       $0.0504$} &  \shortstack{$0.126$ \\       $0.0329$} & \shortstack{$1$ \\       $0.0372$} & \shortstack{$8.5\times 10^{-4}$ \\ $0.0169$} \\
    \hline
    \shortstack{ER \\ $\lambda_2^{(2)}=3.02$} &  \shortstack{$2.8\times 10^{-4}$ \\        $0.0251$ }&  \shortstack{$0.011$\\         $0.0253$}&  \shortstack{$0.0123$ \\        $0.0173$} & \shortstack{$8.5\times 10^{-4}$ \\        $0.0123$} & \shortstack{$1$ \\ $0.0182$}  \\
    \hline
  \end{tabular}
  \caption{Comparison of the overlap (first row in each cell) between the Fiedler modes in each layer, i.e. $({\bm u}_2^{(1)}\cdot{\bm u}_2^{(2)})^2$ and the amplification factor (second row), i.e. $\frac{\sum_i\langle y_i^2\rangle}{\sum_i\langle x_i^2 \rangle}$, for a single cycle; two Watts-Strogatz network with first- and second-nearest neighbors coupling and rewiring probability $p_{\mathrm{rewiring}}=0.05$ (WS I), $0.35$ (WS II); a Barab\'asi-Albert network with $m=4$ (BA); an Erd\H{o}s-R\'enyi with edge probability $p_{\mathrm{edge}}=0.15$ (ER). All networks have size $n=50$\,.  Red cells: The overlap between ${\bm u}_2^{(1)}$ and ${\bm u}_2^{(2)}$ is small for the pair WS I/WS II. However the sum of the overlaps between $\{{\bm u}_2^{(1)},{\bm u}_3^{(1)}\}$ and $\{{\bm u}_2^{(2)},{\bm u}_3^{(2)}\}$ is about $0.5$ thus explaining the amplification observed.}\label{tab1}
\end{table*}

\begin{figure*}
    \centering
    \includegraphics[scale=0.4]{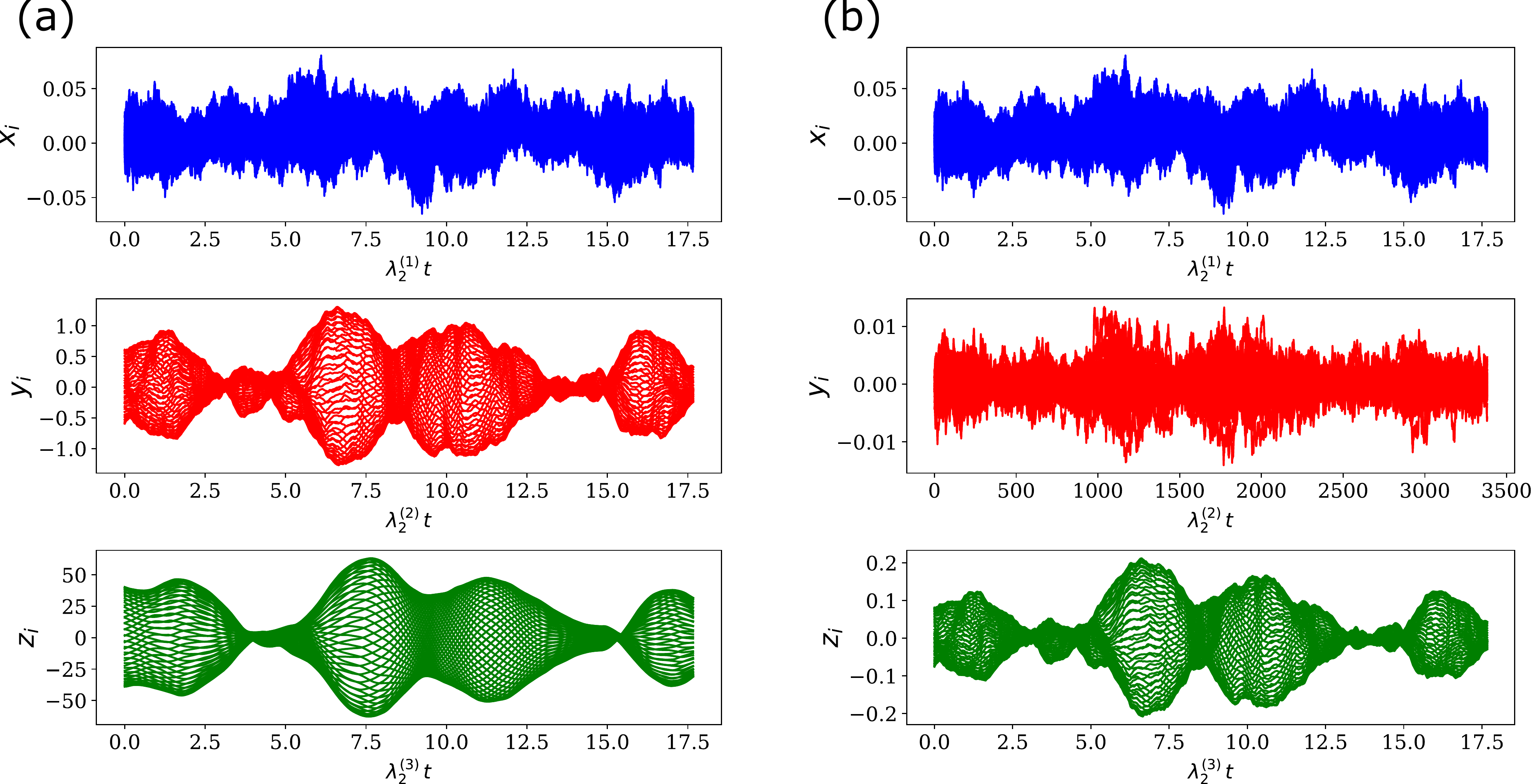}
    \caption{(a)-(b) Three-layer system trajectories $x_i$, $y_i$ and $z_i$ respectively in layer 1, 2 and 3. Amplification of the fluctuations obtained for (a): the same cycle network of size $n=50$ in each of the three layers, i.e. $\mathbb{L}^{(1)}=\mathbb{L}^{(2)}=\mathbb{L}^{(3)}$\,; (b): the same cycle network of size $n=50$ in layers 1 and 3, i.e. $\mathbb{L}^{(1)}=\mathbb{L}^{(3)}$\,, and the same Erd\H{o}s-R\'enyi network as in Tab.~\ref{tab1} for layer 2.}
    \label{fig2}
\end{figure*}

\textbf{More than two layers}.-- So far I considered two-layer systems. However, it is interesting to go one step further and have more than two layers. For example, one may add a third layer that is influenced by the second one exactly as the latter is influenced by the first one. The dynamics of that third layer would then read,
\begin{eqnarray}
\dot{z}_i &=& - \sum_{j=1}^n \mathbb{L}_{ij}^{(3)}\,z_j + f_i(\{y_k\},\{z_k\}) \quad  i=1,...n\,,\label{eq8}
\end{eqnarray}
where $f_i$ is the inter-layer coupling function defined above.
Even though it is possible to calculate analytically the second moment of the $z_i$'s, here I only show numerical results. I illustrate two interesting effects in the case of three layers. First, the amplification of the fluctuations obtained in Fig.~\ref{fig1}(b),(c) for the same network with low algebraic connectivity in each layer becomes even stronger when a third layer of the same network is added. Indeed, Fig.~\ref{fig2}(a) shows the case of three layers of the same cycle network with nearest neighbors. Fluctuations in the third layers are strongly amplified compared to the first one. Significant amplification factors might then be achieved simply by adding layers of the same network with low algebraic connectivity. Second, amplification of the fluctuations in the third layer is still possible even if they were reduced in the second one. The latter is shown in Fig.~\ref{fig2}(b) where the first and the third layer are cycle networks with nearest neighbors and the second layer is an Erd\H{o}s-R\'enyi network with a higher algebraic connectivity. Due to the only marginal overlap between $\bm u_{2}^{(1)}$ and $\bm u_{2}^{(2)}$\,(see Tab.\ref{tab1}), fluctuations are reduced in the second layer. However, in the third layer, fluctuations are still amplified compared to the first one. This phenomenon can be understood as follows. As the second layer reacts faster than the first one, i.e. $\lambda_\alpha^{(2)}\gg \lambda_\alpha^{(1)}$\,, $\alpha\ge 2$\,, it is able to follow closely the slowest modes of the first layer and therefore, to transmit them to the third layer. Then, as the third layer has the same coupling network with low algebraic connectivity as the first layer, fluctuations may still be amplified.\\

\textbf{More general coupling function}.-- The results obtained so far are for a simple choice of coupling function given above Eq.~(\ref{eq3}). Here I briefly show that similar effects are possible for a more general coupling function that reads,
\begin{eqnarray}
f_i(\{x_k\},\{y_k\}) =  - (\mu{y}_i - \nu\overline{x}_i) \,,
\end{eqnarray}
where $\overline{x}_i=x_i - n^{-1}\sum_j x_j$ and $\mu,\nu\in \mathbb{R}$\,. Following the same steps as for the previous coupling function one eventually obtains for the second moment in the second layer,
\begin{widetext}
\begin{align}\label{eq52}
\begin{split}
\langle y_i^2 \rangle = \frac{\nu^2\eta_0^2}{2}\sum_{\alpha, \beta, \gamma} \sum_{k,l} \frac{{u_{\gamma,k}^{(1)}}u_{\gamma,l}^{(1)} {u_{\alpha,k}^{(2)}}u_{\beta,l}^{(2)} [2\lambda_\gamma^{(1)} +\lambda_\alpha^{(2)} + \lambda_\beta^{(2)} + 2\mu] }{\lambda_\gamma^{(1)} (\lambda_\alpha^{(2)} + \lambda_\beta^{(2)} + 2\mu) ( \lambda_\gamma^{(1)} + \lambda_\alpha^{(2)} + \mu) ( \lambda_\gamma^{(1)} + \lambda_\beta^{(2)} + \mu)  }{u_{\alpha,i}^{(2)}}u_{\beta,i}^{(2)} \,.
\end{split}
\end{align}
\end{widetext}
Then, assuming that the two layers have the same coupling network eventually leads to,
\begin{align}\label{eq62}
\begin{split}
\langle y_i^2 \rangle = \frac{\nu^2\eta_0^2}{2}\sum_{\alpha}  \frac{u_{\alpha,i}^2}{\lambda_\alpha (\lambda_\alpha+\mu)(2\lambda_\alpha +\mu)} \,.
\end{split}
\end{align}
First, for $\mu=0$, $\nu=1$ one correctly recovers Eq.~(\ref{eq6}). Second, if $\nu=1$, $\mu>0$, the amplification effect obtained for the simpler coupling function is still possible as long as $|2\lambda_2 +\mu|<1$\,. Third, quite interestingly, for $\nu=-1$, $\mu<0$ and $|\lambda_2+\mu|<1$, i.e. repulsive coupling, significant amplification might occur. Apart from these three cases, one can ensure that no amplification happens by simply choosing $\mu>1-\lambda_2$\,. The more complicated case of two different networks is left to further investigations.\\

\textbf{Colored noise}.-- The amplification phenomena described in previous sections was derived analytically for $\eta_i$'s being white-noises, i.e. uncorrelated in time, and spatially independent. However, it is not limited to white-noise and might be even more important for colored noise, i.e. correlated in time. For example, one may take noisy sequences $\eta_i$'s such that $\langle \eta_i(t)\eta_j(t')  \rangle = \delta_{ij}\,\eta_0^2\,e^{-|t-t'|/\tau_0}$\,, where $\tau_0$ is the typical correlation time. The latter are standardly obtained from an Orstein-Uhlenbeck process. Even though the calculations are doable, the expressions are then complicated and not very insightful. I therefore only discuss the limit of long correlation time, i.e. $\lambda_2^{(1)}\tau_0$, $\lambda_2^{(2)}\tau_0\ge 1$\,. In such a situation the second moment in the first and second layers are given by,
\begin{eqnarray}
\langle x_i^2 \rangle &=& {\eta_0^2}\sum_\alpha \frac{{u_{\alpha,i}^{(1)}}^2}{{\lambda_\alpha^{(1)}}^2}\,,\label{eqcol1}\\
\langle y_i^2 \rangle &=&{\eta_0^2}\sum_{\alpha, \beta, \gamma} \sum_{k,l} \frac{{u_{\gamma,k}^{(1)}}u_{\gamma,l}^{(1)} {u_{\alpha,k}^{(2)}}u_{\beta,l}^{(2)}}{{\lambda_\gamma^{(1)}}^2 \lambda_\alpha^{(2)}\lambda_\beta^{(2)}}{u_{\alpha,i}^{(2)}}u_{\beta,i}^{(2)} \,.\label{eqcol2}
\end{eqnarray}
Unlike for the white-noise case, here one clearly remarks what happens, even for two different networks in the two layers. Given that the overlap between $\bm u_{\gamma}^{(1)}$ and the slowest modes $\bm u_{\beta}^{(2)}$ is finite and not too small, fluctuation can be strongly amplified by having a second layer with $|\lambda_\alpha^{(2)}|\ll1$\,. Comparing Eqs.~(\ref{eqcol1}), (\ref{eqcol2}) to Eqs.~(\ref{eq4}), (\ref{eq5}), one concludes that colored noise with a long correlation time leads to stronger amplification or reduction of the fluctuations than white-noise.
\\

\textbf{Conclusion}.-- Using a modal decomposition of a multi-layer system, I showed that noise originally injected in one layer might be significantly amplified when transmitted to other connected layers. In particular, the amplification strongly depends on the overlap between the eigenmodes of the network Laplacian matrices in each layer and their corresponding eigenvalues. On the one hand, when the overlap is finite and the algebraic connectivity low, fluctuations are amplified in the second layer. On the other hand, if the overlap is marginal or if the algebraic connectivity is high, then fluctuations are likely to be reduced in the second layer. 
Moreover, important amplification factors can be achieved in multi-layer systems by simply adding layers with the same poorly connected network. In addition to that, amplification of the fluctuations might still happen even when they are reduced at intermediate layers. As shown above, fluctuation amplification might be even more significant in case of colored noise with correlation times longer than the intrinsic system's time-scales. These results highlight how inter-dependent systems might be vulnerable to noisy signals. Ways to correct such vulnerabilities would be either to make sure that the algebraic connectivity in each layer is large enough or that the eigenmodes of connected layers do not overlap. The latter should be considered in future research. The amplification effect described here might also be useful to develop inference algorithms~\cite{Xue19}.

Interestingly, the results presented in this letter give indications about how noise input should be correlated in space and time in order to possibly induce larger or smaller fluctuations in single layer diffusively coupled systems. Extension of this work should consider the latter fact into more details, as well as different inter-layer coupling functions. Also, investigating moments higher than the second~\cite{Tyl22} or including inertia for the dynamical agents to potentially uncover resonances in case of time-dependent inter-layer coupling~\cite{Bau20} represent two research avenues of interest.  \\

This work has been supported by the Laboratory Directed Research and Development program of Los Alamos National Laboratory under project number 20220797PRD2.

%


\begin{thebibliography}{28}
\bibitem [{rem({\natexlab{b}})}]{rem2}%
  \BibitemOpen
  \href@noop {} {\bibfield  {journal} {\bibinfo  {journal} {A more general
  coupling function is considered below}\ } ({\natexlab{b}})}\BibitemShut
  {NoStop}%
\end{thebibliography}

\end{document}